\documentclass{cernrep}

\usepackage[strings]{underscore}
\usepackage[T1]{fontenc}

\renewcommand{\rmd}{{\rm d}}

\newcommand{\be}{\begin{equation}}
\newcommand{\ee}{\end{equation}}
\newcommand{\bes}{\begin{equation*}}
\newcommand{\ees}{\end{equation*}}
\newcommand{\bea}{\begin{eqnarray}}
\newcommand{\eea}{\end{eqnarray}}
\newcommand{\cmcub}{cm$^{-3}$}
\newcommand{\fhalf}{\frac{1}{2}}
\newcommand{\half}{{1/2}}
\newcommand{\eps}{\varepsilon}
\newcommand{\downbox}[1]{\ensuremath{_{\rm #1}}}
\newcommand{\Wcm}{W\,cm$^{-2}$}
\newcommand{\ddt}[1]{\frac{\rmd #1}{\rmd t}}
\newcommand{\mum}{$\mu$m}
\newcommand{\dbyd}[2]{\frac{\partial #1}{\partial #2}}
\newcommand{\Tstrut}{{}}
\newcommand{\Bstrut}{{}}

\begin{document}
\title{Introduction to Plasma Physics}
\author{P. Gibbon}
\institute{Forschungszentrum J\"ulich GmbH, Institute for Advanced Simulation, J\"ulich Supercomputing Centre, J\"ulich, Germany}
\maketitle

\begin{abstract}
These notes are intended to provide a brief primer in plasma physics, introducing common definitions, basic properties, and typical processes found in plasmas. These concepts are inherent in contemporary plasma-based accelerator schemes, and thus provide a foundation for the more advanced expositions that follow in this volume. No prior knowledge of plasma physics is required, but the reader is assumed to be familiar with basic electrodynamics and fluid mechanics.
\end{abstract}

\section{Plasma types and definitions}

Plasmas are often described as the fourth state of matter, alongside gases, liquids and solids,  a definition which
does little to illuminate their main physical attributes. In fact, a plasma can exhibit behaviour characteristic of all three
of the more familiar states, depending on its density and temperature, so we obviously need to look for other distinguishing features.  A simple textbook definition of a plasma \cite{chen:book,dendy:book} would be: a \textit{quasi-neutral} gas of charged particles showing \textit{collective} behaviour. This may seem precise enough, but the rather fuzzy-sounding terms of `quasi-neutrality' and `collectivity' require further explanation.  The first of these, `quasi-neutrality', is actually just a mathematical way of saying that even though the  particles making up a plasma consist of free electrons and ions, their overall charge densities cancel each other in equilibrium. So if $n_{\rm e}$ and $n_{\rm i}$ are, respectively, the number densities of electrons and ions with charge state $Z$,
then these are \textit{locally balanced}, i.e.
\begin{equation}
n_{\rm e} \simeq Z n_{\rm i}.
\label{quasineut}
\end{equation}

The second property, `collective' behaviour, arises because of the long-range nature of the $1/r$ Coulomb potential, which means that local disturbances in equilibrium can have a strong influence on remote regions of the plasma.
In other words, macroscopic fields usually dominate over short-lived microscopic fluctuations, and a net charge imbalance $\rho=e(Zn_{\rm i}-n_{\rm e})$ will immediately give rise to an electrostatic field according to Gauss's law,
\[
\nabla\cdot \mathbf{E} = \rho/\varepsilon_0.
\]
Likewise,  the same set of charges moving with velocities $v_{\rm e}$ and $v_{\rm i}$ will give rise to a \textit{current} density $J=e(Zn_{\rm i}v_{\rm i}-n_{\rm e}v_{\rm e})$. This in turn induces a magnetic field according to Amp\`eres law,
\[
\nabla \times \mathbf{B} = \mu_0 \mathbf{J}.
\]
It is these internally driven electric and magnetic fields that largely determine the dynamics of the plasma, including its response to externally applied fields through particle or laser beams---as, for example, in the case of plasma-based accelerator schemes.

Now that we have established what plasmas are, it is natural to ask where we can find them.  In fact they are rather ubiquitous: in the cosmos, 99\% of the visible universe---including stars, the interstellar medium and jets of material from various astrophysical objects---is in a plasma state. Closer to home, the ionosphere, extending from around 50~km (equivalent to 10 Earth radii) to 1000~km, provides vital protection from solar radiation to life on Earth. Terrestrial plasmas can be found in fusion devices (machines designed to confine, ignite and ultimately extract energy from deuterium--tritium fuel), street lighting, industrial plasma torches and etching processes, and lightning discharges. Needless to say, plasmas play a central role in the topic of the present school, supplying the medium to support very large travelling-wave field structures for the purpose of accelerating particles to high energies. Table~\ref{lpps} gives a brief overview of these various plasma types and their properties.

\begin{table}[h]
\begin{center}
\caption{Densities and temperatures of various plasma types}
\begin{tabular}{lcc}
\hline\hline
    \textbf{Type} \hspace{2cm}   & \textbf{Electron density}   & \textbf{ Temperature}  \\
  &  \textbf{$n_{\rm e}$ (\cmcub)}         & \textbf{ $T_{\rm e}$ (eV$^\mathrm{a}$)} \\ \hline
Stars    &  $10^{26}$      &  $2\times 10^{3}$ \\
Laser fusion    &  $10^{25}$      &  $3\times 10^{3}$ \\
Magnetic fusion     &  $10^{15}$     & $10^{3}$ \\
Laser-produced  &  $10^{18}$--$10^{24}$      &  $10^{2}$--$10^{3}$\\
Discharges  &  $ 10^{12}$       &  1--10 \\
Ionosphere & $ 10^{6}$       &  0.1 \\
Interstellar medium &  1       &  $10^{-2}$ \\
\hline \hline
$^\mathrm{a}$ 1~eV $\equiv$ 11\,600~K.
\end{tabular}
\end{center}
\label{lpps}
\end{table}

\subsection{Debye shielding} \label{Debye_sheath}

\begin{figure}[ht]
\begin{center}
\includegraphics[totalheight=1.6in]{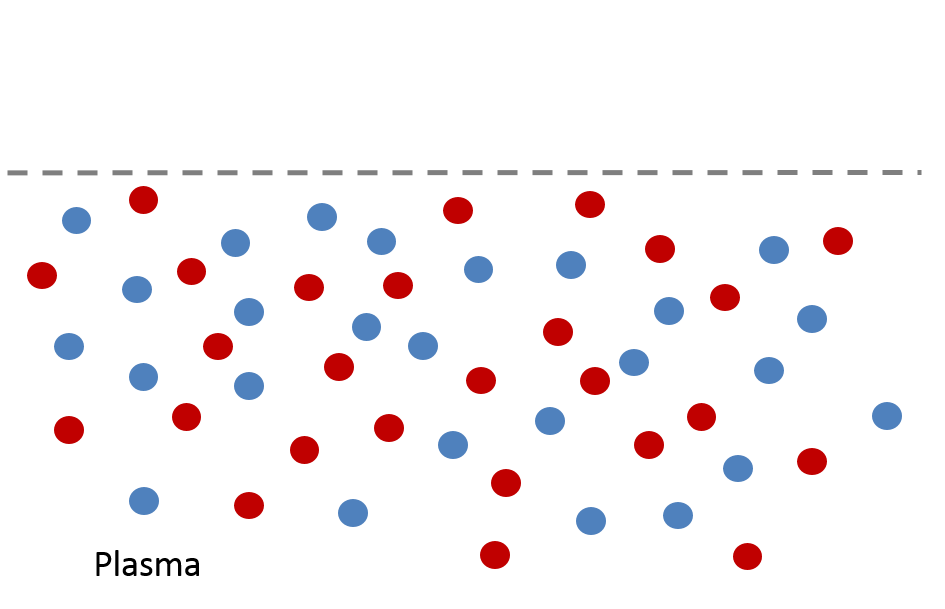}
 \hspace{0.5cm}\includegraphics[totalheight=1.5in]{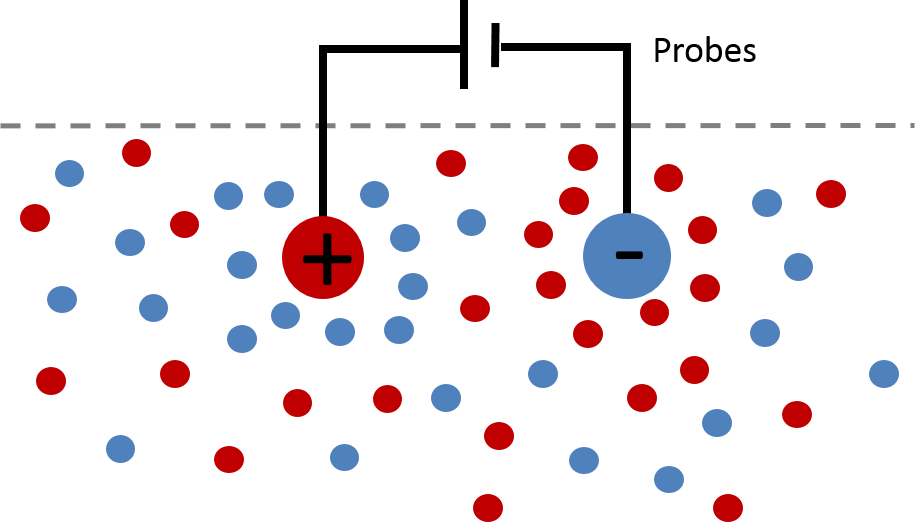}\\
 \caption{Debye shielding of charged spheres immersed in a plasma \label{probes} }
\end{center}
\end{figure}

In most types of plasma, quasi-neutrality is not just an ideal equilibrium state; it is a state that
the plasma actively tries to achieve by readjusting the local charge distribution in response to a disturbance. Consider a hypothetical experiment in which a positively charged ball is immersed in a plasma; see Fig.~\ref{probes}. After some time, the ions in the ball's vicinity will be repelled and the electrons will be attracted, leading to an altered average charge density in this region. It turns out that we can calculate the potential $\phi(r)$ of this ball after such a readjustment has taken place.

First of all, we need to know how fast the electrons and ions actually move. For equal ion and electron temperatures ($T_{\rm e}=T_{\rm i}$), we have
\be
\fhalf m_{\rm e} \overline{v}_{e}^2 = \fhalf m_{\rm i} \overline{v}_{i}^2 = \frac{3}{2} k_{\rm B}T_{\rm e}.
\ee
Therefore, for a hydrogen plasma, where $Z=A=1$,
$$
\frac{\overline{v}_{\rm i}}{\overline{v}_{\rm e}}=\left(\frac{m_{\rm e}}{m_{\rm i}}\right)^{\half} = \left(\frac{m_{\rm e}}{Am_{\rm p}}\right)^{\half} = \frac{1}{43}.
$$
In other words, the ions are almost stationary on the electron time-scale. To a good approximation, we  often write
\be
n_{\rm i} \simeq n_0,
\label{ions_n0}
\ee
where $n_0=N_{\rm A}\rho_{\rm m}/A$ is the material (e.g.\ gas) number density, with $\rho_{\rm m}$ being the usual mass density and $N_{\rm A}$ the Avogadro constant.  In thermal equilibrium, the electron density follows a Boltzmann distribution \cite{chen:book},
\be
n_{\rm e} = n_{\rm i}\exp(e\phi/k_{\rm B}T_{\rm e}),
\label{Boltzmann}
\ee
where $n_{\rm i}$ is the ion density, $k_{\rm B}$ is the Boltzmann constant, and $\phi(r)$ is the potential created by the external disturbance.  From Gauss's law (Poisson's equation), we can also write
\be
\nabla^2\phi = -\frac{\rho}{\eps_0}=-\frac{e}{\eps_0}(n_{\rm i}-n_{\rm e}).
\label{Poisson}
\ee
So now we can combine (\ref{Poisson}) with (\ref{Boltzmann}) and (\ref{ions_n0}) in spherical geometry\footnote{~$\nabla^2\rightarrow\dfrac{1}{r^2}\dfrac{\rmd}{\rmd r}\Bigl({r}^2\dfrac{\rmd\phi}{\rmd r}\Bigr)$.} to eliminate $n_{\rm e}$ and arrive at a physically meaningful solution:
\be
\phi_{\rm D} = \frac{1}{4\pi\eps_0 }\frac{\exp(-r/\lambda_{\rm D})}{r}.
\label{pot-debye}
\ee
This condition supposes that $\phi\rightarrow 0$ at $r=\infty$.  The characteristic length-scale $\lambda_{\rm D}$ inside the exponential factor is known as the \textit{Debye length}, and is given by
\be
\lambda_{\rm D} = \left(\frac{\eps_0k_{\rm B}T_{\rm e}}{e^2n_{\rm e}}\right)^\half
= 743 \left(\frac{T_{\rm e}}{\mbox{eV}}\right)^\half\biggl(\frac{n_{\rm e}}{\mbox{\cmcub}}\biggr)^{-\half} \mbox{cm} .
\label{lambdaD}
\ee
The Debye length is a fundamental property of nearly all plasmas of interest, and depends equally on the plasma's temperature and density. An \textit{ideal} plasma has many particles per Debye sphere, i.e.
\be
N_{\rm D} \equiv n_{\rm e}\frac{4\pi}{3}\lambda_{\rm D}^3 \gg 1,
\label{N_{\rm D}}
\ee
which is a prerequisite for the collective behaviour discussed earlier. An alternative way of expressing this condition is via the so-called \textit{plasma parameter},
\be
g \equiv \frac{1}{n_{\rm e}\lambda_{\rm D}^3},
\label{pl_{\rm p}arameter}
\ee
which is essentially the reciprocal of $N_{\rm D}$. Classical plasma theory is based on the assumption that $g\ll 1$, which implies dominance of collective effects over collisions between particles. Therefore, before we refine our plasma classification, it is worth taking a quick look at the nature of collisions between plasma particles.

\subsection{Collisions in plasmas}
Where $N_{\rm D} \leq 1$, screening effects are reduced and collisions will dominate the particle dynamics.  In intermediate regimes, collisionality is usually measured via the \textit{electron--ion collision rate}, given by
\be
\nu_{\rm ei} = \displaystyle \frac{\pi^{3/2}n_{\rm e} Ze^4\ln\Lambda}{2^{1/2}(4\pi\varepsilon_0)^2m_{\rm e}^2v_{\rm te}^3}\;\mbox{s}^{-1},
\label{nu_{\rm e}i}
\ee
where $v_{\rm te}\equiv \sqrt{k_{\rm B}T_{\rm e}/m_{\rm e}}$ is the electron thermal velocity and $\ln\Lambda$ is a slowly varying term, called the Coulomb logarithm, which typically takes a numerical value of order 10--20.  The numerical coefficient in expression (\ref{nu_{\rm e}i}) may vary between different texts depending on the definition used. Our definition is consistent with that in Refs.\ \cite{kruer:book} and \cite{huba:book}, which define the collision rate according to the average time taken for a thermal electron to be deflected by $90^\circ$ via multiple scatterings from fixed ions. The collision frequency can also be written as
\[
\frac{\nu_{\rm ei}}{\omega_{\rm p}} \simeq \frac{Z\ln\Lambda}{10 N_{\rm D}}\quad \mbox{with } \ln\Lambda\simeq 9N_{\rm D}/Z,
\]
where $\omega_{\rm p}$ is the electron plasma frequency defined below in Eq.~(\ref{omega-p}).

\subsection{Plasma classification}
\begin{figure}[ht]
\begin{center}
\includegraphics[totalheight=3.0in]{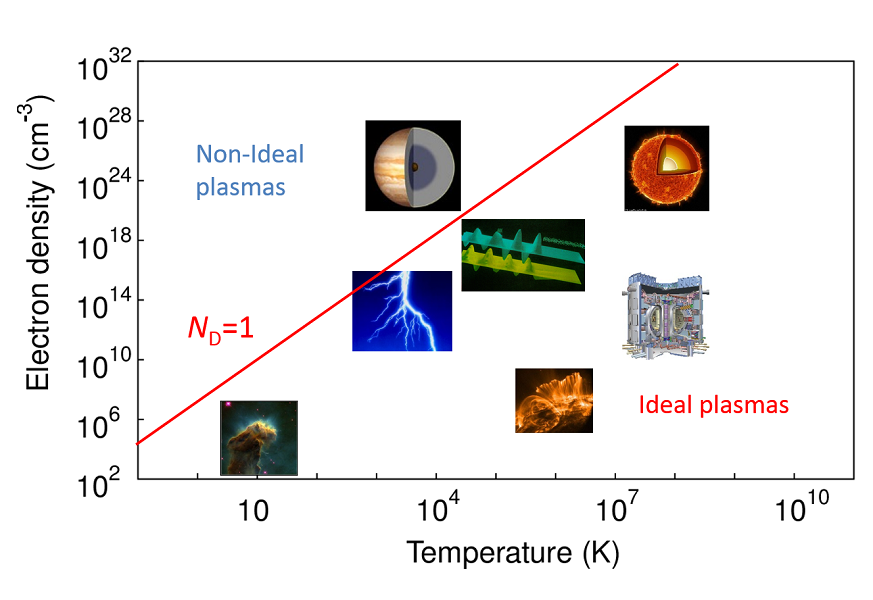}\\
\caption{Examples of plasma types in the density--temperature plane \label{den-temp2}}
\end{center}
\end{figure}

Armed with our definition of plasma ideality, Eq.\ \eqref{N_{\rm D}}, we can proceed to make a classification of plasma types in density--temperature space. This is illustrated for a few examples in Fig.~\ref{den-temp2}; the `accelerator' plasmas of interest in the present school are found in the middle of this chart, having densities corresponding to roughly atmospheric pressure and temperatures of a few eV ($10^4$~K) as a result of field ionization; see Section~\ref{field_{\rm i}oniz}.

\subsection{Plasma oscillations}
\begin{figure}[t]
\begin{center}
\includegraphics[totalheight=2in]{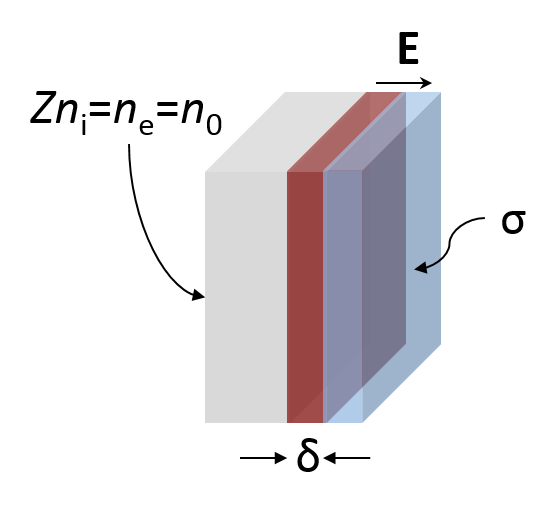}\\
\caption{Slab or capacitor model of an oscillating electron layer \label{Langmuir-osc}}
\end{center}
\end{figure}

So far we have considered characteristics, such as density and temperature, of a plasma in equilibrium. We can also ask how fast the plasma will respond to an external disturbance, which could be due to electromagnetic waves (e.g.\ a laser pulse) or particle beams.  Consider a quasi-neutral plasma slab in which an electron layer is displaced from its initial position by a distance $\delta $, as illustrated in Fig.~\ref{Langmuir-osc}. This creates two `capacitor' plates with surface charge $\sigma = \pm en_{\rm e}\delta $, resulting in an electric field
\[
\mathbf{E} = \frac{\sigma}{\eps_0} = \frac{en_{\rm e}\delta }{\eps_0}.
\]
The electron layer is accelerated back towards the slab by this restoring force according to
\[
m_{\rm e} \frac{\rmd v}{\rmd t} = -m_{\rm e} \frac{\rmd^2\delta }{\rmd t^2} = -eE = \frac{e^2n_{\rm e}\delta }{\eps_0},
\]
or
 \[
 \frac{\rmd^2\delta }{\rmd t^2} + \omega_{\rm p}^2\,\delta  = 0
 \] where
\be
\omega_{\rm p} \equiv \left(\frac{e^2n_{\rm e}}{\varepsilon_0m_{\rm e}}\right)^{\half} \simeq 5.6\times 10^4\biggl(\frac{n_{\rm e}}{\mbox{\cmcub}}\biggr)^\half \;\mbox{s}^{-1}
\label{omega-p}
\ee
is the \textit{electron plasma frequency}.

This quantity can be obtained via another route by returning to the Debye sheath problem of Section~\ref{Debye_sheath} and asking how quickly it would take the plasma to adjust to the insertion of the foreign charge.
For a plasma of temperature $T_{\rm e}$, the response time to recover quasi-neutrality is just the ratio of the Debye length to the thermal velocity $v_{\rm te}\equiv \sqrt{k_{\rm B}T_{\rm e}/m_{\rm e}}$; that is,
$$
t_{\rm D} \simeq \frac{\lambda_{\rm D}}{v_{\rm te}} =  \left(\frac{\varepsilon_0k_{\rm B}T_{\rm e}}{e^2n_{\rm e}}\cdot\frac{m}{k_{\rm B}T_{\rm e}}\right)^\half
= \omega_{\rm p}^{-1}.
$$
 \begin{figure}[h]
\begin{center}
\includegraphics[totalheight=2in]{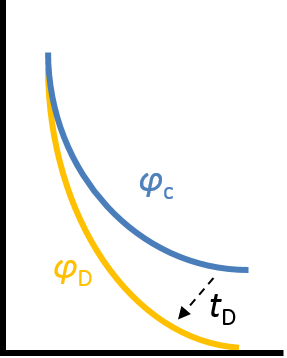}\hspace{1cm}\includegraphics[totalheight=1.8in]{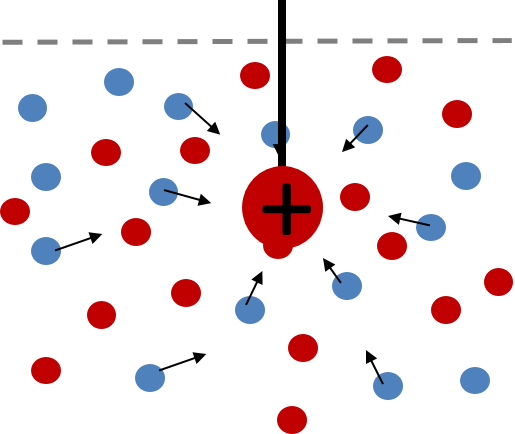}
\end{center}
\caption{Response time to form a Debye sheath}
\end{figure}

If the plasma response time is shorter than the period of a external electromagnetic field (such as a laser), then this radiation will be \textit{shielded out}.  To make this statement more quantitative, consider the ratio
$$
\frac{\omega_{\rm p}^2}{\omega^2} = \frac{e^2n_{\rm e}}{\varepsilon_0m_{\rm e}}\cdot \frac{\lambda^2}{4\pi^2c^2}.
$$
Setting this to unity defines the wavelength $\lambda_\mu$ for which $n_{\rm e}=n_{\rm c}$, or
\be
n_{\rm c} \simeq 10^{21}\lambda_\mu^{-2}\;\mbox{\cmcub}.
\label{n-c}
\ee
Radiation with wavelength $\lambda>\lambda_\mu$ will be reflected. In the pre-satellite/cable era, this property was exploited to good effect in the transmission of long-wave radio signals, which utilizes reflection from the ionosphere to extend the  range of reception.

Typical gas jets have $P\sim 1$~bar and $n_{\rm e} = 10^{18}$--$10^{19}$~\cmcub, and  the critical density for a glass laser is $n_{\rm c}(1 \mu) = 10^{21}$~\cmcub.  Gas-jet plasmas are therefore \textit{underdense}, since $\omega^2/\omega_{\rm p}^2 = n_{\rm e}/n_{\rm c} \ll 1$.  In this case, \textit{collective effects} are important if $\omega_{\rm p} \tau\downbox{int} > 1$, where $\tau\downbox{int}$ is some characteristic interaction time, such as the duration of a laser pulse or particle beam entering the plasma. For example, if $\tau\downbox{int} = 100$~fs and $n_{\rm e} = 10^{17}$~\cmcub, then $\omega_{\rm p} \tau\downbox{int}= 1.8$ and we will need to consider the plasma response on the interaction time-scale.  Generally this is the situation we seek to exploit in all kinds of plasma applications, including short-wavelength radiation, nonlinear refractive properties, generation of  high electric/magnetic fields and, of course, particle acceleration.
\begin{figure}[h]
\begin{center}
\begin{minipage}[t]{0.35\textwidth}
(a)\\
\includegraphics[totalheight=1.4in]{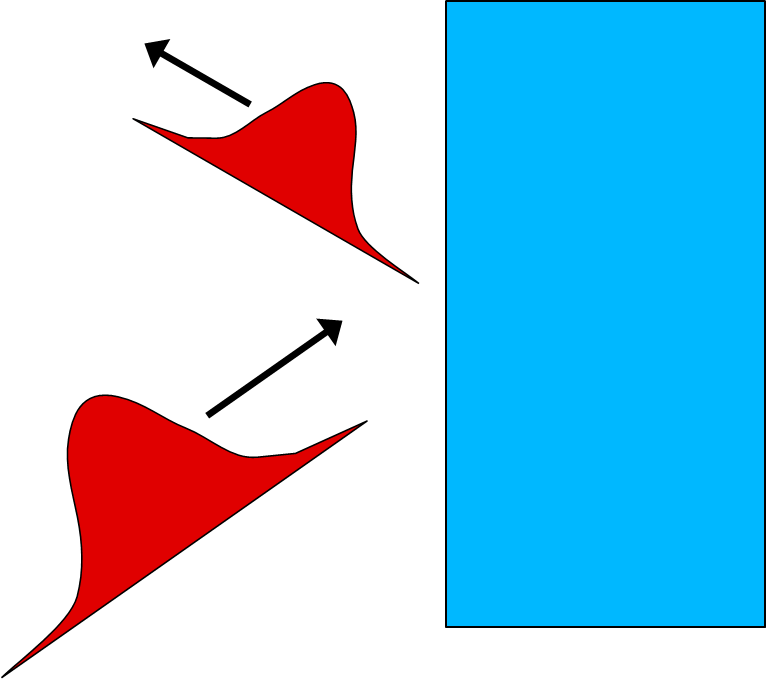}
\end{minipage}  \hspace{1cm}
\begin{minipage}[t]{0.35\textwidth}
(b)\\
\includegraphics[totalheight=1.3in]{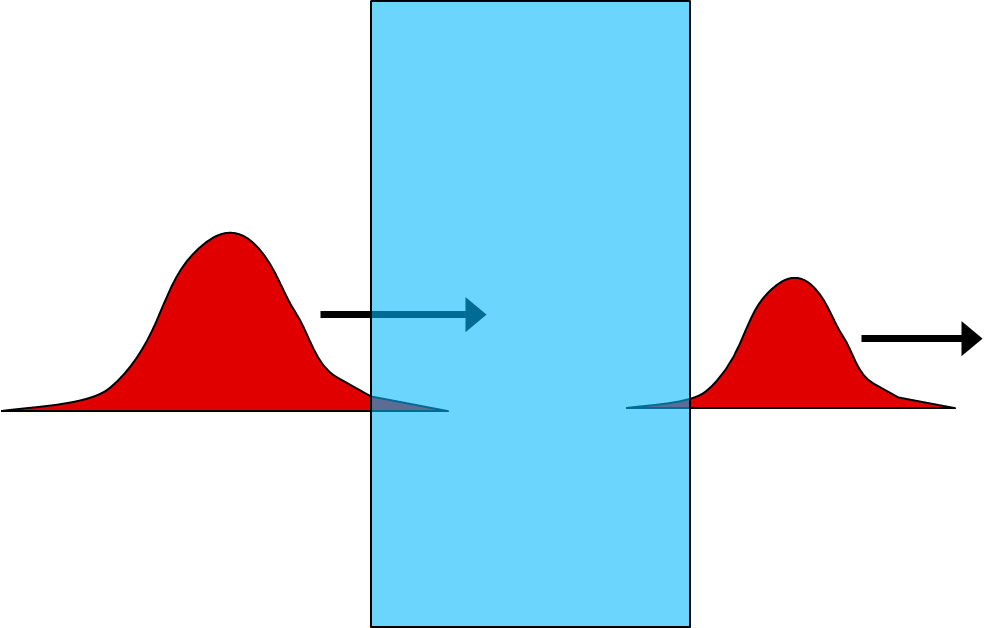}
\end{minipage}
\caption{(a)~Overdense plasma, with $\omega<\omega_{\rm p}$, showing mirror-like behaviour. (b)~Underdense plasma, with $\omega>\omega_{\rm p}$, which behaves like a nonlinear refractive medium. \label{odense_udense}}
\end{center}
\end{figure}

\subsection{Plasma creation} \label{field_{\rm i}oniz}
Plasmas are created via ionization, which can occur in several ways: through collisions of fast particles with atoms; through photoionization by electromagnetic radiation; or via electrical breakdown in strong electric fields.
The latter two are examples of \textit{field ionization}, which is the mechanism most relevant to the plasma accelerator context. To get some idea of when field ionization occurs, we need to know the typical field strength required to strip electrons away from an atom.  At the Bohr radius\index{Bohr~radius}
\bes
a_{\rm B} = \frac{\hbar^2}{me^2} = 5.3\times 10^{-9} \; \mbox{cm},
\ees
the electric field strength is
\be
E_{\rm a}    = \frac{e}{4\pi\varepsilon_0a_{\rm B}^2} 
   \simeq 5.1\times 10^9 \; \mbox{V\,m}^{-1}.
\ee
This threshold can be expressed as the so-called \textit{atomic intensity},
\be
I_{\rm a}   = \displaystyle\frac{\varepsilon_0 c E_{\rm a}^2 }{2} 
    \label{I-atomic}
    \simeq \displaystyle 3.51\times 10^{16} \; \mbox{\Wcm}.
\ee
A laser intensity of $I_{\rm L} > I_{\rm a}$ will therefore guarantee ionization for any target
material, though in fact  ionization can occur well below this threshold (e.g.\ around $10^{14}$ \Wcm\ for hydrogen) due to \textit{multiphoton} effects.  Simultaneous field ionization of many atoms produces a plasma with electron density $n_{\rm e}$ and temperature $T_{\rm e} \sim 1$--$10$~eV.

\subsection{Relativistic threshold}

Before we discuss wave propagation in plasmas, it is useful to have some idea of the strength of the external fields used to excite them. To do this, we consider the classical equation of motion for an electron exposed to a linearly polarized laser field $\mathbf{E} =\hat{y}E_0\sin\omega t$:
\bea
\ddt{v} \simeq \frac{-eE_0}{m_{\rm e}} \sin\omega t. \nonumber
\eea
This implies that the electron will acquire a velocity
\be
 v = \dfrac{eE_0}{m_{\rm e}\omega} \cos\omega t = v\downbox{osc} \cos\omega t,
\ee
which is usually expressed in terms of a dimensionless oscillation amplitude
\be
\large
a_0 \equiv \frac{v\downbox{osc}}{c} \equiv \frac{p\downbox{osc}}{m_{\rm e}c} \equiv \frac{eE_0}{m_{\rm e}\omega c}.
\label{vos}
\ee
In many articles and books $a_0$ is referred to as the `quiver' velocity or momentum; it can exceed unity, in which case the normalized momentum (third expression) is more appropriate, since the real particle velocity is just pinned to the speed of light.
The laser intensity $I_{\rm L}$ and wavelength $\lambda_{\rm L}$ are related to $E_0$ and $\omega$ through
\[
I_{\rm L} = \fhalf \varepsilon_0cE_0^2, \qquad \lambda_{\rm L} = \frac{2\pi c}{\omega}.
\]
By substituting these into (\ref{vos}) one can show that
\be
a_0 \simeq 0.85 (I\downbox{18}\lambda_\mu^2)^\half,
\label{a0}
\ee
where
\[
I\downbox{18}=\frac{I_{\rm L}}{10^{18}\:\mbox{\Wcm}}, \qquad \lambda_\mu = \frac{\lambda_{\rm L}}{\mu {\rm m}}.
\]
From this expression it can be seen that we will have relativistic electron velocities,
 or $a_0 \sim 1$, for intensities $I_{\rm L}\geq 10^{18}$~\Wcm, at wavelengths $\lambda_{\rm L} \simeq 1$~\mum.

\section{Wave propagation in plasmas}

The theory of wave propagation is an important subject in its own right, and has inspired a vast body of literature and a number of textbooks \cite{kruer:book,dougherty:chapter,boyd:book}.  There are a great many possible ways in which plasmas can support waves, depending on the local conditions, the presence of external electric and magnetic fields, and so on.
Here we will concentrate on two main wave forms: longitudinal oscillations of the kind we have encountered already, and electromagnetic waves.  To derive and analyse wave phenomena, there are several possible theoretical approaches, with the suitability of each depending on the length- and time-scales of interest, which in laboratory plasmas can range from nanometres to metres and from femtoseconds to seconds. These approaches are:
\begin{enumerate}
\item[(i)] first-principles $N$-body molecular dynamics;
\item[(ii)]  phase-space methods---the Vlasov--Boltzmann equation;
\item[(iii)]  two-fluid equations;
\item[(iv)]  magnetohydrodynamics (single magnetized fluid).
\end{enumerate}
The first is rather costly and limited to much smaller regions of plasma than usually needed to describe the common types of wave. Indeed, the number of particles needed for first-principles modelling of a tokamak would be around $10^{21}$; a laser-heated gas requires $10^{20}$ particles, still way out of reach of even the most powerful computers available. Clearly a more tractable model is needed, and in fact many plasma phenomena can be analysed by assuming that each charged particle component  of  density $n_s$ and velocity $\mathbf{u}_s$ behaves in a fluid-like manner, interacting with other species ($s$) via the electric and magnetic fields; this is the idea behind approach (iii). The rigorous way to derive the governing equations in this approximation is via \textit{kinetic theory}, starting from method (ii) \cite{kruer:book,dendy:book}, which is beyond the scope of this paper. Finally, slow wave phenomena on more macroscopic, ion time-scales can be handled with approach (iv) \cite{dendy:book}.

For the present purposes, we therefore start from the two-fluid equations for a plasma with finite temperature ($T_{\rm e} > 0$) that is assumed to be collisionless ($\nu\downbox{ie}\simeq 0$) and non-relativistic, so that the fluid velocities are such that $u \ll c$.  The equations governing the plasma dynamics under these conditions are
\begin{align}
&\dbyd{n_s}{t} + \nabla\cdot(n_s\mathbf{u}_s) = 0, \label{continuity}\\
&n_sm_s\ddt{\mathbf{u}_s} = n_sq_s(\mathbf{E} + \mathbf{u}_s\times \mathbf{B}) - \nabla P_s, \label{momentum-density}\\
&\ddt{}(P_sn_s^{-\gamma_s}) = 0, \label{energy-density}
\end{align}
where $P_s$ is the thermal pressure of species $s$ and $\gamma_s$ the specific heat ratio, or $(2+N)/N$ with $N$ the number of degrees of freedom.

The continuity equation (\ref{continuity}) tells us that (in the absence of ionization or recombination) the number of particles \textit{of each species} is conserved.  Noting that the charge and current densities can be written as $\rho_s = q_sn_s$ and  $\mathbf{J}_s = q_sn_s\mathbf{u}_s$, respectively, Eq.~(\ref{continuity}) can be rewritten as
\be
\dbyd{\rho_s}{t} + \nabla\cdot\mathbf{J}_s = 0, \label{charge}
\ee
which expresses the conservation of \textit{charge}.

Equation (\ref{momentum-density}) governs the motion of a fluid element of species $s$ in the presence of electric and magnetic fields $\mathbf{E}$ and $\mathbf{B}$.   In the absence of fields, and assuming strict quasi-neutrality ($n_{\rm e}=Zn_{\rm i}=n;\ \,\mathbf{u}_{\rm e}=\mathbf{u}_{\rm i}=\mathbf{u}$), we recover the more familiar \textit{Navier--Stokes} equations
\begin{align}
\dbyd{\rho}{t} + \nabla\cdot(\rho\mathbf{u}) &= 0, \nonumber\\[-7pt]
\label{navier_stokes} \\[-7pt]
\dbyd{\mathbf{u}}{t} + (\mathbf{u}\cdot\nabla)\mathbf{u} &= \frac{1}{\rho}\nabla P. \nonumber
\end{align}
By contrast, in the plasma accelerator context we usually deal with time-scales over which the ions can be assumed to be motionless, i.e.\ $\mathbf{u}_{\rm i}=0$, and also unmagnetized plasmas, so that the momentum equation reads
\be
n_{\rm e}m_{\rm e}\ddt{\mathbf{u}_{\rm e}} = -e\, n_{\rm e}\mathbf{E}  - \nabla P_{\rm e}.
\label{electron_mom}
\ee
Note that $\mathbf{E}$ can include both external and internal  field components (via charge separation).

\subsection{Longitudinal (Langmuir) waves}

\begin{figure}[ht]
\begin{center}
\includegraphics[totalheight=2.2in]{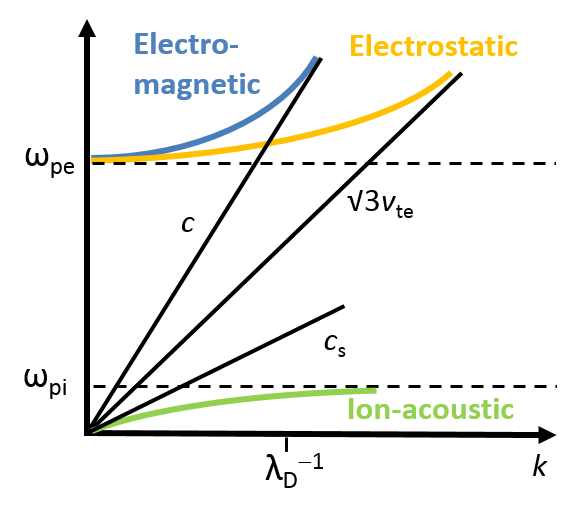}
\end{center}
\caption{Schematic illustration of dispersion relations for Langmuir, electromagnetic and ion-acoustic waves \label{disp_curves}}
\end{figure}

A characteristic property of plasmas is their ability to transfer momentum and energy via collective motion. One of the most important examples of this is the oscillation of electrons against a stationary ion background, or \textit{Langmuir waves}.  Returning to the two-fluid model, we can simplify (\ref{continuity})--(\ref{energy-density}) by setting $\mathbf{u}_{\rm i}=0$, restricting the electron motion to one dimension ($x$) and taking $\dbyd{}{y}=\dbyd{}{z}=0$:
\begin{align}
\dbyd{n_{\rm e}}{t}+\dbyd{}{x}(n_{\rm e}u_{\rm e}) &=0,\nonumber \\
n_{\rm e}\left(\dbyd{u_{\rm e}}{t}+ u_{\rm e}\dbyd{u_{\rm e}}{x}\right) &=  -\frac{e}{m}n_{\rm e}E - \frac{1}{m}\dbyd{P_{\rm e}}{x},\label{Langmuir_1d} \\
\ddt{}\left(\frac{P_{\rm e}}{n_{\rm e}^{\gamma_{\rm e}}}\right) &=0. \nonumber
\end{align}
The system (\ref{Langmuir_1d}) has three equations and four unknowns.
To close it, we need an expression for the electric field, which, since  $\mathbf{B}=0$, can be found from Gauss's law (Poisson's equation) with $Zn_{\rm i}=n_0$:
\be
\dbyd{E}{x} = \frac{e}{\eps_0}(n_0-n_{\rm e}) .
\label{Poisson_1d}
\ee
The system of equations (\ref{Langmuir_1d})--(\ref{Poisson_1d}) is nonlinear and, apart from a few special cases, cannot be solved exactly.  A common technique for analysing waves in plasmas is to \textit{linearize} the equations, which involves assuming that the perturbed amplitudes are small compared to the equilibrium values, i.e.
\begin{align}
n_{\rm e} &= n_0 + n_1, \nonumber \\
u_{\rm e} &= u_1, \nonumber \\
P_{\rm e} &= P_0 + P_1, \nonumber \\
E &= E_1, \nonumber
\end{align}
where $n_1\ll n_0$ and $P_1\ll P_0$.  Upon substituting these expressions  into (\ref{Langmuir_1d})--(\ref{Poisson_1d}) and neglecting all products of perturbations such as $n_1\partial_t u_1$ and $u_1\partial_xu_1$, we get a set of linear equations for the perturbed quantities:
\begin{align}
\dbyd{n_1}{t}+n_0\dbyd{u_1}{x} &= 0,\nonumber \\
\Tstrut n_0\dbyd{u_1}{t} &= -\frac{e}{m}n_0E_1 - \frac{1}{m}\dbyd{P_1}{x}, \label{fluid_lin}\\
\Tstrut \dbyd{E_1}{x} &= -\frac{e}{\eps_0}n_1, \nonumber \\
\Tstrut P_1 &= 3k_{\rm B}T_{\rm e}\,n_1. \nonumber
\end{align}
The expression for $P_1$ results from the specific heat ratio $\gamma_{\rm e}=3$ and from assuming isothermal background electrons, $P_0=k_{\rm B}T_{\rm e}n_0$ (ideal gas); see Kruer's book \cite{kruer:book}.  We can now eliminate $E_1, P_1$ and $u_1$ from (\ref{fluid_lin}) to get
\be
\left(\dbyd{^2}{t^2} - 3v_{\rm te}^2\dbyd{^2}{x^2}+\omega_{\rm p}^2\right)n_1 = 0,
\label{langmuir_wave}
\ee
with $v_{\rm te}^2=k_{\rm B}T_{\rm e}/m_{\rm e}$ and $\omega_{\rm p}$ given by (\ref{omega-p}) as before.  Finally, we look for plane-wave solutions of the form $A = A_0\exp\{\mathrm{i}(\omega t-kx)\}$, so that our derivative operators are transformed as follows: $\dbyd{}{t}\rightarrow \mathrm{i}\omega$ and $\dbyd{}{x}\rightarrow -\mathrm{i}k$.  Substitution into (\ref{langmuir_wave}) yields the Bohm--Gross dispersion relation
\be
\omega^2 = \omega_{\rm p}^2 + 3k^2v_{\rm te}^2.
\label{Bohm-Gross}
\ee
This and other dispersion relations are often depicted graphically on a chart such as that in Fig.~\ref{disp_curves}, which gives an overview of which propagation modes are permitted for low- and high-wavelength limits.

\subsection{Transverse waves}
To describe \textit{transverse} electromagnetic (EM) waves, we need two additional Maxwell's equations, Faraday's law  and Amp\`ere's law, which we will introduce properly later; see Eqs.~(\ref{faraday}) and (\ref{ampere}).  For the time being, it is helpful to simplify things by making use of our previous analysis of  small-amplitude longitudinal waves. Therefore, we linearize and again apply the harmonic approximation $\dbyd{}{t}\rightarrow \mathrm{i}\omega$ to get
\begin{align}
\nabla \times \mathbf{E}_1 & =  - \mathrm{i}\omega\mathbf{B}_1,
\label{faraday_lin}\\
\nabla \times \mathbf{B}_1 & =  \mu_0 \mathbf{J}_1 +\mathrm{i}\eps_0\mu_0\omega\mathbf{E}_1,
\label{ampere_lin}
\end{align}
where the transverse current density is given by
\be
\mathbf{J}_1 = -n_0e\mathbf{u}_1.
\label{current_lin}
\ee
This time we look for pure EM plane-wave solutions with $\mathbf{E}_1 \perp \mathbf{k}$ (see Fig.~\ref{geom_{\rm e}m}) and also assume that the group and phase velocities are  large enough,  $v_{\rm p}, v_g \gg v_{\rm te}$, so that we have a \textit{cold} plasma with $P_{\rm e}=n_0k_{\rm B}T_{\rm e}\simeq 0$.
\begin{figure}[h]
\begin{center}
\includegraphics[totalheight=1.0in]{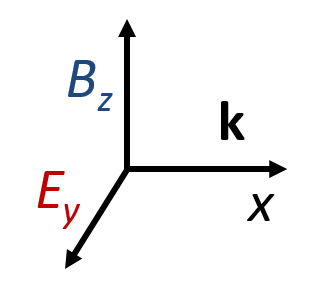}
\end{center}
\caption{Geometry for electromagnetic plane-wave analysis
\label{geom_{\rm e}m}}
\end{figure}

\begin{figure}[ht]
\begin{center}
\includegraphics[totalheight=2.0in]{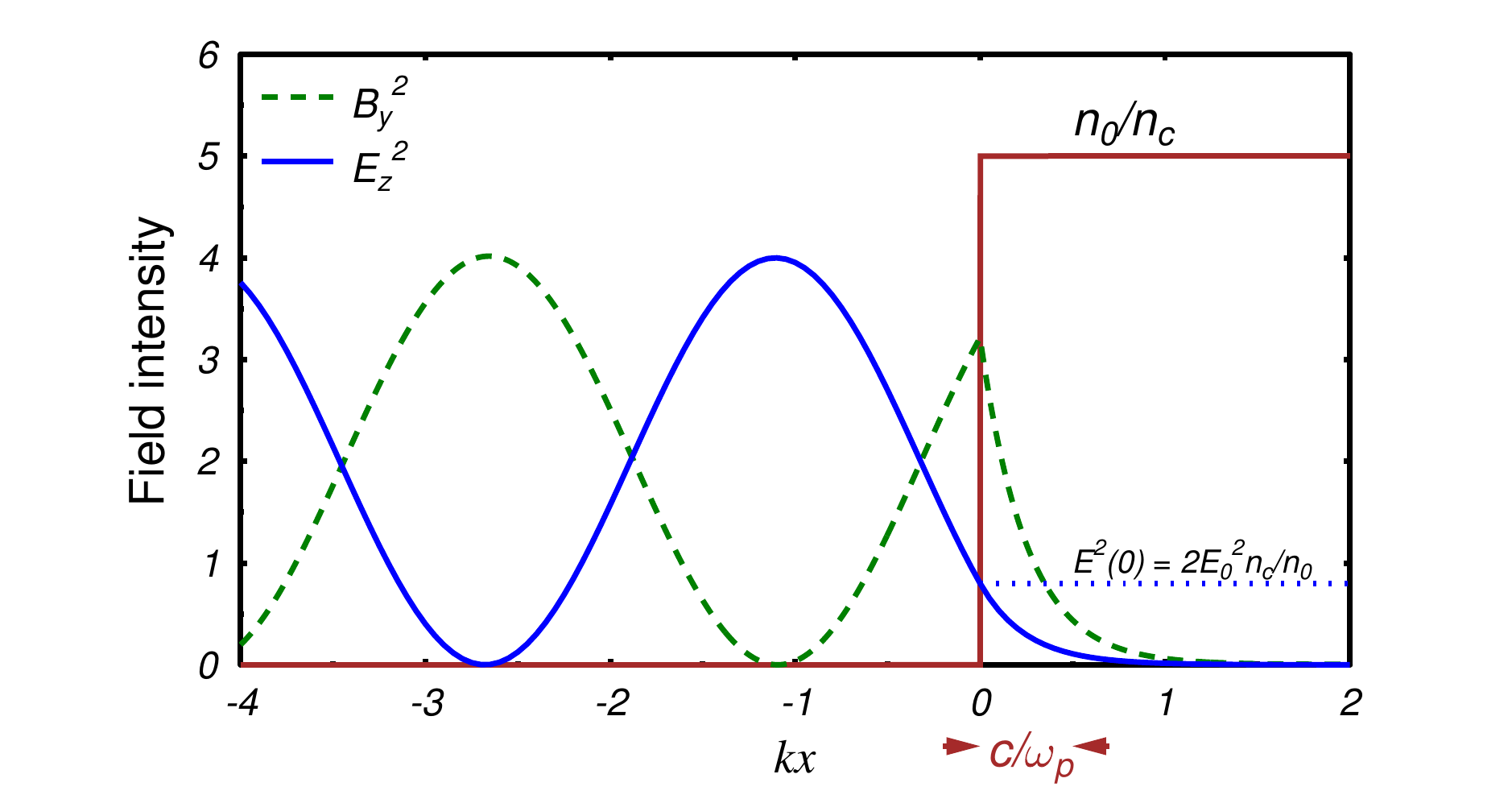}
\end{center}
\caption{Electromagnetic fields resulting from reflection of an incoming wave by an overdense plasma slab \label{skin_fields}}
\end{figure}

The linearized electron fluid velocity and corresponding current are then
\begin{align}
\mathbf{u}_1 &= -\frac{e}{\mathrm{i} \omega m_{\rm e}}\mathbf{E}_1, \nonumber \\[-7pt]
\label{ac-conductivity}\\[-7pt]
\mathbf{J}_1 &= \frac{n_0e^2}{\mathrm{i} \omega m_{\rm e}} \mathbf{E}_1 \equiv \sigma \mathbf{E}_1 ,\nonumber
\end{align}
where $\sigma$ is the \textit{AC electrical conductivity}.  By analogy with dielectric media (see, e.g., Ref.\ \cite{jackson:book}), in which Amp\`ere's law is usually written as $\nabla \times \mathbf{B}_1  = \mu_0 \partial_t{\mathbf{D}}_1$, by substituting (\ref{ac-conductivity}) into (\ref{ampere}) one can show that
\[
\mathbf{D}_1 = \eps_0\varepsilon\mathbf{E}_1,
\]
with
\be
\varepsilon = 1 + \frac{\sigma}{\mathrm{i}\omega\eps_0} = 1-\frac{\omega_{\rm p}^2}{\omega^2}.
\label{Dielectric_function}
\ee
From (\ref{Dielectric_function}) it follows immediately that
\be
\eta \equiv \sqrt{\varepsilon} = \frac{ck}{\omega} = \left(1-\frac{\omega_{\rm p}^2}{\omega^2}\right)^{\half}, \label{refractive_{\rm i}ndex}
\ee
with
\be
\omega^2 = \omega_{\rm p}^2 + c^2k^2.
\label{em_disprel}
\ee
The above expression can also be found directly by elimination of $\mathbf{J}_1$ and $\mathbf{B}_1$ from Eqs.\ (\ref{faraday_lin})--(\ref{ac-conductivity}).  From the dispersion relation (\ref{em_disprel}), also depicted in Fig.~\ref{disp_curves}, a number of important features of EM wave propagation in plasmas can be deduced. For \textit{underdense} plasmas ($n_{\rm e}\ll n_{\rm c}$),
\begin{align}
\mbox{phase velocity} \quad v_{\rm p} &= \frac{\omega}{k} \simeq c\left(1+\frac{\omega_{\rm p}^2}{2\omega^2}\right) > c \,;\nonumber \\
\mbox{group velocity} \quad v_{\rm g} &= \dbyd{\omega}{k} \simeq c\left(1-\frac{\omega_{\rm p}^2}{2\omega^2}\right) < c\, .\nonumber
\end{align}
In the opposite case of an \textit{overdense} plasma, where $n_{\rm e} > n_{\rm c}$, the refractive index $\eta$ becomes imaginary and the wave can no longer propagate, becoming evanescent instead, with a decay length determined by the \textit{collisionless skin depth} $c/\omega_{\rm p}$; see Fig.~\ref{skin_fields}.

\subsection{Nonlinear wave propagation}
So far we have considered purely longitudinal or transverse waves; linearizing the wave equations ensures that any nonlinearities or coupling between these two modes is excluded. While this is a reasonable approximation for low-amplitude waves, it is inadequate for decscribing strongly driven waves in the relativistic regime of interest in plasma accelerator schemes. The starting point of most analyses of nonlinear wave propagation phenomena is the Lorentz equation of motion for the electrons in a \textit{cold} ($T_{\rm e}=0$) unmagnetized plasma, together with Maxwell's equations \cite{kruer:book,gibbon:book}.  We make two further assumptions: (i)~that the ions are initially  singly charged ($Z=1$) and are treated as an immobile ($v_{\rm i}=0$), homogeneous background with $n_0=Zn_{\rm i}$; (ii)~that thermal motion can be neglected, since the temperature remains low compared to the typical oscillation energy in the laser field ($v\downbox{osc}\gg v_{\rm te}$).  The starting equations (in SI units) are then as follows:
\begin{align}
\frac{\partial \mathbf{p}}{\partial t} + (\mathbf{v} \cdot \nabla)\mathbf{p} & =  -e(\mathbf{E} + \mathbf{v }\times \mathbf{B}),
\label{lorentz} \\
\nabla \cdot \mathbf{E} & =   \frac{e}{\varepsilon_0}(n_0 - n_{\rm e}),
\label{poisson}\\
\nabla \times \mathbf{E} & =  - \frac{\partial\mathbf{B}}{\partial t},
\label{faraday}\\
c^2\nabla \times \mathbf{B} & =  -\frac{e}{\varepsilon_0}n_{\rm e}\mathbf{v} +  \frac{\partial\mathbf{E}}{\partial t},
\label{ampere} \\
\nabla \cdot \mathbf{B} & =   0,       \label{divb}
\end{align}
where $\mathbf{p} = \gamma m_{\rm e} \mathbf{v}$ and $\gamma = (1+p^2/m_{\rm e}^2c^2)^\half$.

To simplify matters, we first assume a plane-wave geometry like that in Fig.~\ref{geom_{\rm e}m}, with the transverse electromagnetic fields given by $\mathbf{E}_{\rm L}=(0,E_y,0)$ and $\mathbf{B}_{\rm L}=(0,0,B_z)$.
From Eq.~(\ref{lorentz}), the transverse electron momentum is then simply
\be
p_y = eA_y,
\label{canmom}
\ee
where $E_y~=~\partial A_y/\partial t$.  This relation expresses conservation of canonical momentum.
Substituting $\mathbf{E}=-\nabla\phi-\partial \mathbf{A}/\partial t$ and $\mathbf{B}=\nabla\times\mathbf{A}$ into Amp\`ere's equation (\ref{ampere}) yields
\[
c^2\nabla\times (\nabla\times \mathbf{A}) + \frac{\partial^2 \mathbf{A}}{\partial t^2}  =
\frac{\mathbf{J}}{\eps_0} - \nabla\dbyd{\phi}{t},
\]
where the current is given by $\mathbf{J} = -en_{\rm e}\mathbf{v}$.  Now we use a bit of vectorial wizardry, splitting the current into rotational (solenoidal) and irrotational (longitudinal) parts,
\[
\mathbf{J} = \mathbf{J}_\perp + \mathbf{J}_{\parallel} = \nabla\times\mathbf{\Pi} + \nabla{\Psi},
\]
from which we can deduce (see Jackson's book  \cite{jackson:book}) that
\[
\mathbf{J}_{\parallel} - \frac{1}{c^2}\nabla\dbyd{\phi}{t}=0.
\]
Finally, by applying the Coulomb gauge $\nabla\cdot\mathbf{A}=0$ and $v_y=eA_y/\gamma$ from (\ref{canmom}), we obtain
\be
\frac{\partial^2 A_y}{\partial t^2} - c^2\nabla^2 A_y    = \mu_0J_y = -\frac{e^2n_{\rm e}}{\eps_0 m_{\rm e}\gamma} A_y.
\label{emwave}
\ee
The nonlinear source term on the right-hand side contains two important bits of physics: $ n_{\rm e} = n_0 + \delta n$, which couples the EM wave to plasma waves, and $\gamma=\sqrt{1+\mathbf{p}^2/m_{\rm e}^2c^2}$, which introduces relativistic effects through the increased electron inertia.  Taking the \textit{longitudinal} component of the momentum equation (\ref{lorentz}) gives
\[
\ddt{}(\gamma m_{\rm e}v_x) = -eE_x - \frac{e^2}{2m_{\rm e}\gamma}\dbyd{A_y^2}{x}.
\]
We can eliminate $v_x$ using the $x$ component of Amp\`ere's law (\ref{ampere}):
\[
0=-\frac{e}{\eps_0}n_{\rm e} v_x + \dbyd{E_x}{t}.
\]
And the electron density can be determined via Poisson's equation (\ref{poisson}):
\[
n_{\rm e} = n_0 - \frac{\varepsilon_0}{e}\dbyd{E_x}{x}.
\]

\begin{figure}[ht]
\begin{center}
\includegraphics[totalheight=2.1in]{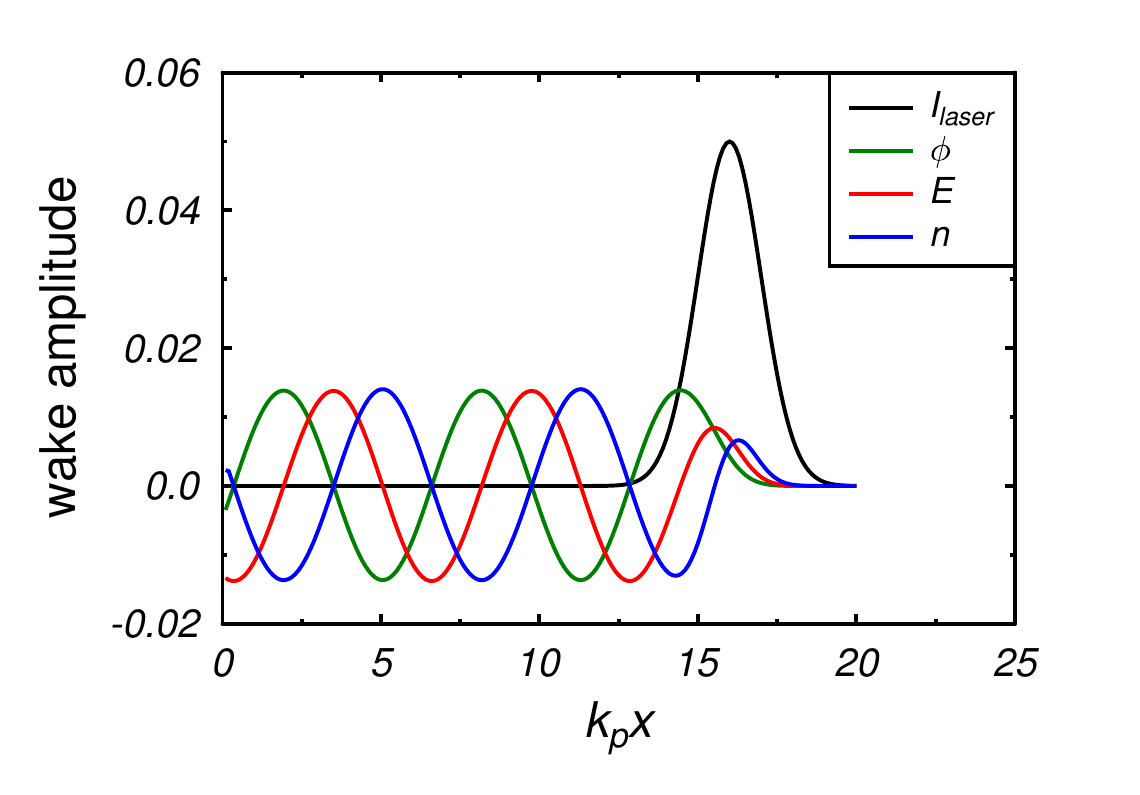}
\includegraphics[totalheight=2.1in]{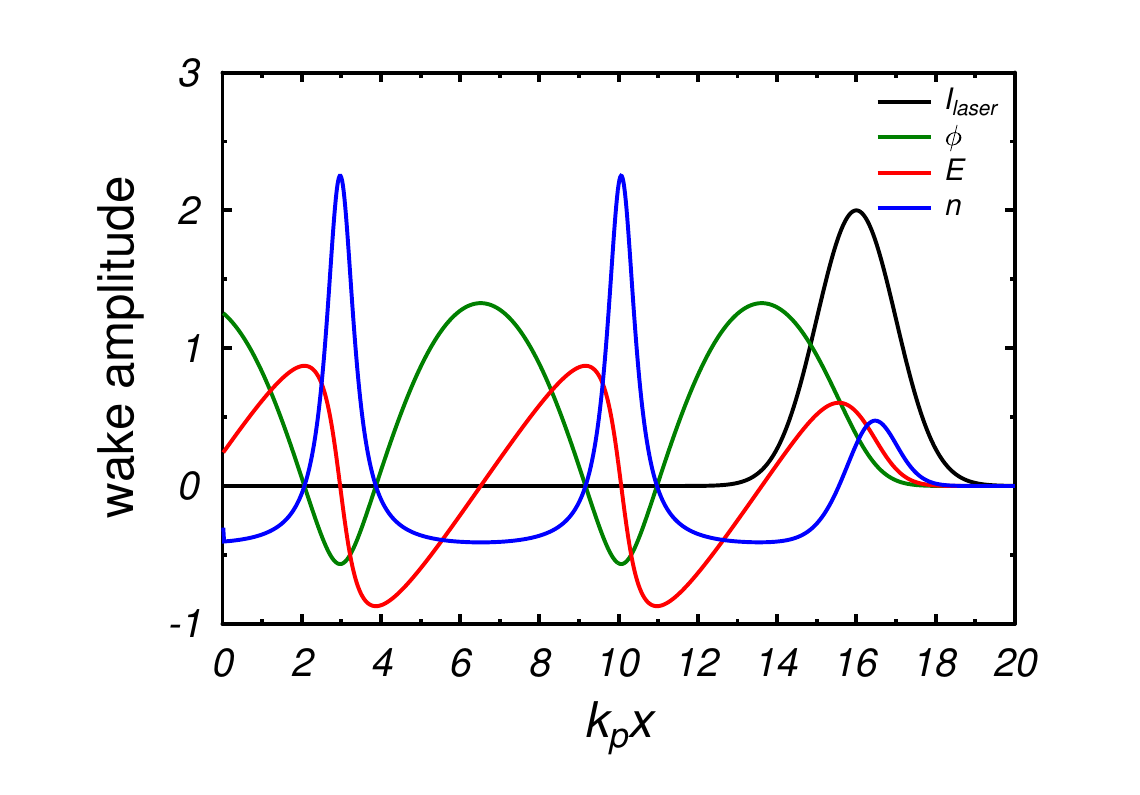}
\caption{Wakefield excitation by a short-pulse laser propagating in the positive $x$ direction in the linear regime (left) and nonlinear regime (right).}
\label{qsa_wake}
\end{center}
\end{figure}

The above (closed) set of equations can in principle be solved numerically for arbitrary pump strengths.  For the moment, we simplify things by linearizing the \textit{plasma} fluid quantities. Let
\begin{align}
n_{\rm e} &\simeq n_0 + n_1 + \cdots ,\nonumber \\
v_x &\simeq v_1 + v_2 + \cdots, \nonumber
\end{align}
and neglect products of perturbations such as $n_1v_1$. This leads to
\be
\left( \dbyd{^2 }{t^2} + \frac{\omega_{\rm p}^2}{\gamma_0}\right) E_x = -\frac{\omega_{\rm p}^2e}{2m_{\rm e}\gamma_0^2}\dbyd{ }{x}A_y^2 .
\label{langmuir}
\ee
The driving term on the right-hand side is the \textit{relativistic ponderomotive force}, with $\gamma_0=(1+a_0^2/2)^\half$.  Some solutions of Eq.~(\ref{langmuir}) are shown in Fig.~\ref{qsa_wake}, for low- and high-intensity laser pulses. The properties of the wakes will be discussed in detail in other lectures, but we can already see some obvious qualitative differences between the linear and nonlinear wave forms; the latter are typically characterized by a spiked density profile, a sawtooth electric field, and a longer wavelength.

The coupled fluid equations (\ref{emwave}) and (\ref{langmuir}) and their fully nonlinear counterparts describe a wide range of nonlinear laser--plasma interaction phenomena, many of which are treated in the later lectures of this school, including plasma wake generation, blow-out regime
laser self-focusing and channelling, parametric instabilities, and harmonic generation.
Plasma-accelerated particle \textit{beams}, on the other hand,  cannot be treated using fluid theory and require a more sophisticated kinetic approach, usually assisted by numerical models solved with the aid of powerful supercomputers.


\newpage
\appendix
\section{Useful constants and formulae}
\label{sec:app}

\begin{table}[h]
\begin{center}
\caption{Commonly used physical constants}
\label{tab:constants}
\begin{tabular}{lccc}
\hline\hline
\textbf{Name} & \textbf{Symbol} & \textbf{Value (SI)} & \textbf{Value (cgs)}\\
\hline
 Boltzmann constant   &   $k_{\rm B}$ & $1.38\times 10^{-23}$~J\,K$^{-1}$ & $1.38\times 10^{-16}$ erg\,K$^{-1}$ \Bstrut \\
Electron charge 	& $e$ 	& $1.6\times 10^{-19}$ C & $4.8\times 10^{-10}$ statcoul \Bstrut\\
Electron mass 	& $m_{\rm e}$	& $9.1\times 10^{-31}$ kg & $9.1\times 10^{-28}$ g \Bstrut\\
Proton mass 		& $m_{\rm p}$	& $1.67\times 10^{-27}$ kg & $1.67\times 10^{-24}$ g \Bstrut\\
Planck constant	& $h$	& $6.63\times 10^{-34}$ J\,s & $6.63\times 10^{-27}$ erg-s \Bstrut\\
Speed of light	& $c$	& $3\times 10^8$ m\,s$^{-1}$ & $3\times 10^{10}$ cm\,s$^{-1}$\Bstrut\\
Dielectric constant	& $\varepsilon_0$ & $8.85\times10^{-12} $ F\,m$^{-1}$ & --- \Bstrut\\
Permeability constant 	& $\mu_0$	& $4\pi\times10^{-7}$ & --- \Bstrut\\
Proton/electron mass ratio	& $m_{\rm p}/m_{\rm e}$ & 1836 & 1836 \Bstrut\\
Temperature = 1eV	& $e/k_{\rm B}$  & 11\,604 K	& 11\,604 K\Bstrut\\
Avogadro number 		& $N_{\rm A}$	   & $6.02\times 10^{23}$ mol$^{-1}$ &$6.02\times 10^{23}$ mol$^{-1}$ \Bstrut\\
Atmospheric pressure & 1 atm	& $1.013\times 10^5$ Pa	& $1.013\times 10^6$ dyne cm$^{-2}$\\
\hline\hline
\end{tabular}
\end{center}
\end{table}

\begin{table}[h]
\begin{center}
\caption{Formulae in SI and cgs units}
\label{tab:formulae}
\begin{tabular}{lccc}
\hline\hline
\textbf{Name} & \textbf{Symbol} & \textbf{Formula (SI)} & \textbf{Formula (cgs)}\\
\hline
Debye length		& $\lambda_{\rm D}$	& $\displaystyle \left(\frac{\varepsilon_0k_{\rm B}T_{\rm e}}{e^2n_{\rm e}}\right)^{1/2}$m
					& $\displaystyle \left(\frac{k_{\rm B}T_{\rm e}}{4\pi e^2n_{\rm e}}\right)^{1/2}$cm \Tstrut\\[10pt]
Particles in Debye sphere & $N_{\rm D}$ & $ \displaystyle \frac{4\pi}{3}\lambda_{\rm D}^3$ & $ \displaystyle\frac{4\pi}{3}\lambda_{\rm D}^3$\\[5pt]
Plasma frequency (electrons)& $\omega_{\rm pe}$  & $\displaystyle \left(\frac{e^2n_{\rm e}}{\varepsilon_0 m_{\rm e}}\right)^{1/2}$s$^{-1}$	& $\displaystyle \left(\frac{4\pi e^2n_{\rm e}}{m_{\rm e}}\right)^{1/2}$s$^{-1}$\\[10pt]
Plasma frequency (ions)  &  $\omega_{\rm pi}$  & $\displaystyle \left(\frac{Z^2e^2n_{\rm i}}{\varepsilon_0 m_{\rm i}}\right)^{1/2}$s$^{-1}$	& $\displaystyle \left(\frac{4\pi Z^2 e^2n_{\rm i}}{m_{\rm i}}\right)^{1/2}$s$^{-1}$\\[10pt]
Thermal velocity & $v_{\rm te}=\omega_{\rm pe}\lambda_{\rm D}$	& $\displaystyle \left(\frac{k_{\rm B}T_{\rm e}}{m_{\rm e}}\right)^{1/2}$m\,s$^{-1}$ & $\displaystyle \left(\frac{k_{\rm B}T_{\rm e}}{m_{\rm e}}\right)^{1/2}$cm\,s$^{-1}$\\[10pt]
Electron gyrofrequency & $\omega_{\rm c}$	& $\displaystyle eB/m_{\rm e}$~s$^{-1}$	& $eB/m_{\rm e}$~s$^{-1}$\\[5pt]
Electron--ion collision frequency & $\nu_{\rm ei}$ & $\displaystyle \frac{\pi^{3/2}n_{\rm e} Ze^4\ln\Lambda}{2^{1/2}(4\pi\varepsilon_0)^2m_{\rm e}^2v_{\rm te}^3}$~s$^{-1}$&  $\displaystyle \frac{4(2\pi)^{1/2}n_{\rm e} Ze^4\ln\Lambda}{3m_{\rm e}^2v_{\rm te}^3}$~s$^{-1}$\\[12pt]
Coulomb logarithm & $\ln\Lambda$	& $\displaystyle\ln \frac{9N_{\rm D}}{Z}$ &$\displaystyle\ln \frac{9N_{\rm D}}{Z}$\\[7pt]
\hline\hline
\end{tabular}
\end{center}
\end{table}

\begin{table}[h]
\begin{center}
\caption{Useful formulae, with $T_{\rm e}$ in eV, $n_{\rm e}$ and $n_{\rm i}$ in cm$^{-3}$, and wavelength $\lambda_{\rm L}$ in $\mu$m}
\label{tab:handy}
\begin{tabular}{ll}
\hline\hline
Plasma frequency & $\displaystyle \omega_{\rm pe} = 5.64\times10^4n_{\rm e}^{1/2}$~s$^{-1}$ \\[7pt]
Critical density & $\displaystyle n_{\rm c} = 10^{21}\lambda\downbox{L}^{-2}$~cm$^{-3}$ \Tstrut \Bstrut\\[5pt]
Debye length & $\displaystyle \lambda_{\rm D} = 743~T_{\rm e}^{1/2}n_{\rm e}^{-1/2}$ cm \Tstrut \\[5pt]
Skin depth & $\displaystyle \delta = c/\omega_{\rm p} = 5.31\times10^5n_{\rm e}^{-{1/2}}$ cm \Tstrut\\[5pt]
Electron--ion collision frequency & $\displaystyle \nu_{\rm ei} = 2.9\times10^{-6}n_{\rm e}T_{\rm e}^{-{3/2}}\ln\Lambda$~\,s$^{-1}$ \Tstrut\\[3pt]
Ion--ion collision frequency & $\displaystyle \nu_{\rm ii} = 4.8\times 10^{-8}Z^4\left(\frac{m_{\rm p}}{m_{\rm i}}\right)^{1/2}n_{\rm i}T_{\rm i}^{-{3/2}}\ln\Lambda $~\,s$^{-1}$ \Tstrut\\[7pt]
Quiver amplitude & $\displaystyle a_0\equiv\frac{p_{\rm osc}}{m_{\rm e}c} = \left(\frac{I\lambda_{\rm L}^2}{1.37\times10^{18}\;\mbox{W\,cm}^{-2}\,\mu \mbox{m}^2}\right)^{1/2}$ \Tstrut\\[7pt]
Relativistic focusing threshold & $\displaystyle P_{c} = 17\left(\frac{n_{\rm c}}{n_{\rm e}}\right)$ GW \Tstrut\\[7pt]
\hline\hline
\end{tabular}
\end{center}
\end{table}

\end{document}